\documentclass[12pt]{article}
\usepackage{amsmath}
\usepackage{color}
\usepackage[colorlinks]{hyperref}

\def\comment#1{}
\input{tcilatex}

\begin{document}

\title{Description of compatible differential-geometric Poisson brackets of
the first order}
\author{Maxim Pavlov}
\date{}
\maketitle
\tableofcontents

\section{Introduction}

One of the most interesting questions of the classical differential geometry
which has appeared at studying of semi-Hamiltonian systems of hydrodynamical
type is the description of the surfaces admitting not trivial deformations
with preservation of principal directions and principal curvatures. Then the
number of essential parameters on which such deformations depend, is
actually equal to number various local Hamiltonian structures of
corresponding system of hydrodynamical type \cite{viniti}. Such local
Hamiltonian structures are determined by differential-geometrical Poisson
brackets of the first order (see \cite{Novik}). In the same paper a
bi-Hamiltonian structure of the system of averaged one-phase solutions of
KdV have been considered. Later multi-Hamiltonian structures of systems of
hydrodynamical type were studied in \cite{Pavlov}, \cite{Maks+Tsar}. The
given work is devoted to the description of bi-Hamiltonian structures in
language of orthogonal curvilinear coordinate nets. The alternative approach
(by the method of an inverse scattering transform) has been used in \cite%
{Mokh}. In \cite{fer} it has been shown, that the description of pairs
compatible differential-geometric Poisson brackets of the first order is
equivalent to the solution of the over-determined system on rotation
coefficients $\beta _{ik}$ of orthogonal curvilinear coordinate nets 
\begin{equation}
\begin{array}{c}
\partial _{i}\beta _{jk}=\beta _{ji}\beta _{ik}\text{, \ \ \ }i\neq j\neq k,
\\ 
\\ 
\partial _{i}\beta _{ik}+\partial _{k}\beta _{ki}+\sum_{m\neq i,k}\beta
_{mi}\beta _{mk}=0\text{, \ \ \ \ }i\neq k, \\ 
\\ 
\eta _{i}\partial _{i}\beta _{ik}+\eta _{k}\partial _{k}\beta _{ki}+\frac{1}{%
2}\eta _{i}^{\prime }\beta _{ik}+\frac{1}{2}\eta _{k}^{\prime }\beta
_{ki}+\sum_{m\neq i,k}\eta _{m}\beta _{mi}\beta _{mk}=0\text{, \ \ \ \ }%
i\neq k,%
\end{array}
\label{1}
\end{equation}%
where $\eta _{i}(r^{i})$ are some functions, $\partial _{k}\equiv \partial
/\partial r^{k}$, $k=1$, $2$, ..., $N$. The general case $\eta
_{i}(r^{i})\neq \func{const}$ and particular case $\eta _{i}(r^{i})=c_{i}=%
\func{const}$ were described in \cite{fer}, where it has been shown, that (%
\ref{1}) is an integrable system. In that specific case $\eta
_{i}(r^{i})=c_{i}=\func{const}$ the system of equations%
\begin{equation}
\begin{array}{c}
\partial _{i}\beta _{jk}=\beta _{ji}\beta _{ik}\text{, \ \ \ }i\neq j\neq k,
\\ 
\\ 
\partial _{i}\beta _{ik}+\partial _{k}\beta _{ki}+\sum_{m\neq i,k}\beta
_{mi}\beta _{mk}=0\text{, \ \ \ \ }i\neq k, \\ 
\\ 
c_{i}\partial _{i}\beta _{ik}+c_{k}\partial _{k}\beta _{ki}+\sum_{m\neq
i,k}c_{m}\beta _{mi}\beta _{mk}=0\text{, \ \ \ \ }i\neq k,%
\end{array}
\label{3}
\end{equation}%
when $N=3$ was reduced to three pairwise commuting non-evolution equations
(see \cite{fer})%
\begin{equation*}
q_{xt}=\func{ch}q\sqrt{1-q_{x}^{2}}\text{, \ \ }q_{yt}=\func{sh}q\sqrt{%
1-q_{y}^{2}},\text{ \ \ \ \ }q_{xy}=-\sqrt{(1-q_{x}^{2})(1-q_{y}^{2})}\text{,%
}
\end{equation*}%
which are nothing but two modified Sin-Gordon equations and degenerated
twice modified Sin-Gordon equation, respectively. In this paper
generalization of this system (i.e. $\eta _{i}(r^{i})\neq \func{const}$) is
presented for bi-Hamiltonian structure where one metric is flat, and the
second one has curvature of co-dimension 1.

\section{Flat case}

Compatibility condition of the linear system in partial derivatives%
\begin{equation}
\partial _{i}H_{k}=\beta _{ik}H_{i}\text{, \ \ \ \ }i\neq k  \label{a}
\end{equation}%
is the same as compatibility condition for conjugated linear system%
\begin{equation}
\partial _{i}\psi _{k}=\beta _{ki}\psi _{i}\text{, \ \ \ \ }i\neq k.
\label{b}
\end{equation}%
This yields the nonlinear multi-dimensional integrable $N$-wave system (see
for instance, \cite{Dubr}, \cite{Orlov})%
\begin{equation}
\partial _{i}\beta _{jk}=\beta _{ji}\beta _{ik}\text{, \ \ \ }i\neq j\neq k.
\label{one}
\end{equation}%
Any pair of solutions of the first linear system $H_{i}^{(1)}$, $H_{i}^{(2)}$
generates hydrodynamic type system integrable by the generalized hodograph
method \cite{Tsar}%
\begin{equation*}
r_{t}^{i}=\frac{H_{i}^{(2)}}{H_{i}^{(1)}}r_{x}^{i}\text{, \ \ \ \ \ }i=1%
\text{, }2\text{, ..., }N,
\end{equation*}%
written in Riemann invariants $r^{i}$. Particular solutions $\psi _{i}^{(k)}$
of the second linear system (\ref{b}) determine densities $a^{k}$ and fluxes 
$c^{k} $ of conservation laws%
\begin{equation*}
a_{t}^{k}=\partial _{x}c^{k}(\mathbf{a}),
\end{equation*}%
where%
\begin{equation*}
da^{k}=\overset{N}{\sum_{m=1}}\psi _{m}^{(k)}H_{m}^{(1)}dr^{m}\text{, \ \ \
\ }dc^{k}=\overset{N}{\sum_{m=1}}\psi _{m}^{(k)}H_{m}^{(2)}dr^{m}.
\end{equation*}%
Existence of the linear differential operator of the first order connecting
solutions of both linear systems (see for instance, \cite{Maks+Tsar})%
\begin{equation}
H_{i}=\partial _{i}\psi _{i}+\sum_{m\neq i}\beta _{mi}\psi _{m}  \label{c}
\end{equation}%
is equivalent to existence of local Hamiltonian structure 
\begin{equation*}
a_{t}^{i}=\partial _{x}[g^{ik}\frac{\partial h}{\partial a^{k}}],
\end{equation*}%
where flat coordinates $a^{k}$ are determined by \textquotedblright
null\textquotedblright\ solutions of the second linear system%
\begin{equation*}
0=\partial _{i}\psi _{i}^{(k)}+\sum_{m\neq i}\beta _{mi}\psi _{m}^{(k)}\text{%
, \ \ \ \ }k=1\text{, }2\text{, ... , }N,
\end{equation*}%
and non-degenerate symmetric metric%
\begin{equation*}
ds^{2}=\overset{N}{\sum_{m=1}}\left( H_{m}^{(1)}dr^{m}\right)
^{2}=g_{jk}da^{j}da^{k},
\end{equation*}%
which is constant in flat coordinates $a^{k}$%
\begin{equation*}
g^{jk}=\overset{N}{\sum_{m=1}}\psi _{m}^{(j)}\psi _{m}^{(k)}=\func{const}.
\end{equation*}%
Existence of such operator (\ref{c}) imposes the restriction on rotation
coefficients of orthogonal curvilinear coordinate nets $\beta _{ik}$%
\begin{equation}
\partial _{i}\beta _{ik}+\partial _{k}\beta _{ki}+\sum_{m\neq i,k}\beta
_{mi}\beta _{mk}=0\text{, \ \ \ \ }i\neq k,  \label{nul}
\end{equation}%
well-known in differential geometry as the Gauss equation, in above case it
fixes the metric of zero curvature.

It is easy to see that $N$-wave system is invariant under the transformation 
$R^{i}=R^{i}(r^{i})$, $\bar{H}_{m}^{(1)}=H_{m}^{(1)}\mu _{i}^{-1/2}(r^{i})$,
where $\mu _{i}(r^{i})$ and $R^{i}(r^{i})$ are arbitrary functions of a
single variable. If Riemann invariants are fixed ($R^{i}(r^{i})\equiv r^{i}$%
), then \textquotedblleft flatness\textquotedblright\ condition (\ref{nul})
is no longer valid, then solutions of the conjugate problem (\ref{b})
transform in accordance with $\bar{\psi}_{i}=\psi _{i}\mu _{i}^{1/2}(r^{i})$%
; if $\mu _{i}(r^{i})=R^{i\prime ^{2}}(r^{i})$, then a metric preserves
\textquotedblleft flatness" 
\begin{equation*}
ds^{2}=\overset{N}{\sum_{m=1}}\left( H_{m}^{(1)}dr^{m}\right) ^{2}=\overset{N%
}{\sum_{m=1}}\left( \bar{H}_{m}^{(1)}dR^{m}\right) ^{2},
\end{equation*}%
and solutions of the conjugate problem (\ref{b}) are the same%
\begin{equation*}
da^{k}=\overset{N}{\sum_{m=1}}\psi _{m}^{(k)}H_{m}^{(1)}dr^{m}=\overset{N}{%
\sum_{m=1}}\bar{\psi}_{m}^{(k)}\bar{H}_{m}^{(1)}dR^{m}.
\end{equation*}%
Thus, infinite number of metrics%
\begin{equation}
ds^{2}=\overset{N}{\sum_{m=1}}\mu _{m}^{-1}(r^{m})\left(
H_{m}^{(1)}dr^{m}\right) ^{2},  \label{met}
\end{equation}%
are connected to each system of hydrodynamic type and only for finite number
of values $\mu _{i}(r^{i})$ (no more than $N+1$, see \cite{Maks+Fer})
the\textquotedblleft flatness\textquotedblright\ condition is satisfied. In
this paper the case is considered, when just two distinct values $\mu
_{i}(r^{i})$ determine flat metrics. Since coefficients of the diagonal
metric $H_{i}^{2}$ are determined up to multiplication on an arbitrary
function of corresponding Riemann invariant, then the first flat metric
always can be chosen so that one of values $\mu _{i}(r^{i})$ can be fixed to
1. Compatible pairs of local Hamiltonian structures determined by such
metrics studied for instance in \cite{Mokh}. Such pairs are compatible if
their arbitrary linear combination also is local Hamiltonian structure.
Thus, the integrable nonlinear system (see \cite{fer})%
\begin{equation}
\bar{\partial}_{i}\bar{\beta}_{jk}=\bar{\beta}_{ji}\bar{\beta}_{ik}\text{, \
\ \ }i\neq j\neq k,  \label{!}
\end{equation}%
\begin{equation*}
(\lambda +R^{i})\bar{\partial}_{i}\bar{\beta}_{ik}+(\lambda +R^{k})\bar{%
\partial}_{k}\bar{\beta}_{ki}+\frac{1}{2}(\bar{\beta}_{ik}+\bar{\beta}%
_{ki})+\sum_{m\neq i,k}(\lambda +R^{m})\bar{\beta}_{mi}\bar{\beta}_{mk}=0,\
\ \ \ i\neq k.
\end{equation*}%
is a compatibility condition of the linear system%
\begin{equation}
\begin{array}{c}
\bar{\partial}_{i}\bar{\psi}_{j}^{(k)}=\bar{\beta}_{ji}\bar{\psi}_{i}^{(k)}%
\text{, \ \ \ \ }i\neq j, \\ 
\\ 
0=(\lambda +R^{i})\bar{\partial}_{i}\bar{\psi}_{i}^{(k)}+\frac{1}{2}\bar{\psi%
}_{i}^{(k)}+\sum_{m\neq i}(\lambda +R^{m})\bar{\beta}_{mi}\bar{\psi}%
_{m}^{(k)}\text{, \ \ \ \ }k=1\text{, }2\text{, ... , }N,%
\end{array}
\label{va}
\end{equation}%
where $\bar{\partial}=\partial /\partial R^{i}$, \ $\bar{\beta}_{ik}(\mathbf{%
R})=\beta _{ik}(\mathbf{r})/R^{i\prime }(r^{i})$, \ $\bar{\psi}%
_{j}^{(k)}=\psi _{j}^{(k)}$, $\lambda $ is an arbitrary constant of the flat
metric $\tilde{g}^{ii}(\mathbf{r})=(\lambda +R^{i})g^{ii}$.

Thus, we have proved following

\textbf{Lemma}: The reduction of the $N$-wave system (\ref{one})%
\begin{equation*}
\begin{array}{c}
\partial _{i}\beta _{ik}+\partial _{k}\beta _{ki}+\sum_{m\neq i,k}\beta
_{mi}\beta _{mk}=0\text{, \ \ \ \ }i\neq k, \\ 
\\ 
\eta _{i}(r^{i})\partial _{i}\beta _{ik}+\eta _{k}(r^{k})\partial _{k}\beta
_{ki}+\frac{1}{2}\eta _{i}^{\prime }(r^{i})\beta _{ik}+\frac{1}{2}\eta
_{k}^{\prime }(r^{k})\beta _{ik}+\sum_{m\neq i,k}\eta _{m}(r^{m})\beta
_{mi}\beta _{mk}=0\text{, \ \ \ \ }i\neq k,%
\end{array}%
\end{equation*}%
where $\eta _{i}(r^{i})$ is a second solution $\mu _{i}(r^{i})$, is resulted
in (\ref{!}) by scaling $H_{i}(r)=\eta _{i}^{\prime }(r^{i})\tilde{H}_{i}(R)$%
, $\partial /\partial r^{i}=\eta _{i}^{\prime }(r^{i})\partial /\partial
R^{i}$, $\eta _{i}(r^{i})=\tilde{\eta}_{i}(R^{i})\equiv R^{i}$, if $\eta
_{i}(r^{i})\neq c_{i}=\func{const}$.

Thus, in general case bi-Hamiltonian structures of hydrodynamic type systems
are described by solutions of the integrable system%
\begin{equation*}
\begin{array}{c}
\partial _{i}\beta _{jk}=\beta _{ji}\beta _{ik}\text{, \ \ \ \ }i\neq j\neq
k, \\ 
\\ 
\partial _{i}\beta _{ik}+\partial _{k}\beta _{ki}+\sum_{m\neq i,k}\beta
_{mi}\beta _{mk}=0\text{, \ \ \ \ }i\neq k, \\ 
\\ 
r^{i}\partial _{i}\beta _{ik}+r^{k}\partial _{k}\beta _{ki}+\frac{1}{2}\beta
_{ik}+\frac{1}{2}\beta _{ki}+\sum_{m\neq i,k}r^{m}\beta _{mi}\beta _{mk}=0%
\text{, \ \ \ \ }i\neq k.%
\end{array}%
\end{equation*}%
Its spectral problem (\ref{va})

\begin{equation*}
\begin{array}{c}
\partial _{i}\psi _{j}^{(k)}=\beta _{ji}\psi _{i}^{(k)}\text{, \ \ \ \ }%
i\neq j, \\ 
\\ 
0=(\lambda +r^{i})\partial _{i}\psi _{i}^{(k)}+\frac{1}{2}\psi
_{i}^{(k)}+\sum_{m\neq i}(\lambda +r^{m})\beta _{mi}\psi _{m}^{(k)}\text{, \
\ \ \ }k=1\text{, }2\text{, ..., }N%
\end{array}%
\end{equation*}%
is nothing but $N$ commuting systems of \textit{ordinary} differential
equations with respect to every independent variable (Riemann invariant $r^i$%
).

\textbf{Remark}: This multi-dimensional spectral problem has a
multi-dimensional analog of Wronskian: $N(N+1)/2$ \textquotedblleft first
integrals\textquotedblright\ are constraints%
\begin{equation}
g^{sn}=\overset{N}{\sum_{m=1}}(\lambda +r^{m})\psi _{m}^{(s)}\psi _{m}^{(n)}=%
\func{const},  \label{metr}
\end{equation}%
which are nothing but metric coefficients in flat coordinates of \textit{%
mixed} local Hamiltonian structure.

Following G. Darboux for further consideration we should determine
\textquotedblleft first integrals\textquotedblright\ of the Gauss equation
(see (\ref{nul})).

\textbf{Definition}: A \textit{scalar} potential of rotation coefficients of
conjugate curvilinear coordinate nets $V$ is a such function, whose mixed
second derivatives are%
\begin{equation*}
V_{ik}=\beta _{ik}\beta _{ki}\text{, \ \ \ \ \ }i\neq k.
\end{equation*}%
\textbf{Definition}: A \textit{vector} potential of rotation coefficients of
conjugate curvilinear coordinate nets $S_{k}$ is a such function, whose
mixed first derivatives are%
\begin{equation*}
\partial _{i}S_{k}=\beta _{ik}\partial _{k}\beta _{ki}\text{, \ \ \ \ \ }%
i\neq k.
\end{equation*}%
G. Darboux have proved (see \cite{Darboux}), that the Gauss equation,
written via rotation coefficients of conjugate curvilinear coordinate nets (%
\ref{nul}), has $N$ \textquotedblleft first integrals\textquotedblright 
\begin{equation}
S_{i}+\frac{1}{2}\sum_{m\neq i}\beta _{mi}^{2}=n_{i}(r^{i}),  \label{gau}
\end{equation}%
where $n_{i}(r^{i})$ are arbitrary functions (\textquotedblleft integration
factors\textquotedblright ). Similar zero curvature condition%
\begin{equation*}
\eta _{i}(r^{i})\partial _{i}\beta _{ik}+\eta _{k}(r^{k})\partial _{k}\beta
_{ki}+\frac{1}{2}\eta _{i}^{\prime }(r^{i})\beta _{ik}+\frac{1}{2}\eta
_{k}^{\prime }(r^{k})\beta _{ik}+\sum_{m\neq i,k}\eta _{m}(r^{m})\beta
_{mi}\beta _{mk}=0\text{, \ \ \ \ }i\neq k
\end{equation*}%
for metric $\eta _{i}(r^{i})g^{ii}$ has $N$ \textquotedblleft first
integrals\textquotedblright 
\begin{equation}
\eta _{i}S_{i}+\frac{1}{2}\eta _{i}^{\prime }V_{i}+\frac{1}{2}\sum_{m\neq
i}\eta _{m}\beta _{mi}^{2}=k_{i}(r^{i}),  \label{gaus}
\end{equation}%
where $k_{i}(r^{i})$ are another arbitrary functions.

\subsection{Particular case, $\protect\eta _{i}(r^{i})=c_{i}=\func{const}$}

This particular case was considered in \cite{fer}. The spectral problem

\begin{equation*}
\begin{array}{c}
\partial _{i}\psi _{j}^{(k)}=\beta _{ji}\psi _{i}^{(k)}\text{, \ \ \ \ }%
i\neq j, \\ 
\\ 
0=(\lambda +c_{i})\partial _{i}\psi _{i}^{(k)}+\sum_{m\neq i}(\lambda
+c_{m})\beta _{mi}\psi _{m}^{(k)}\text{, \ \ \ \ }k=1\text{, }2\text{, ... , 
}N,%
\end{array}%
\end{equation*}%
determines the nonlinear system (\ref{3})%
\begin{equation*}
\begin{array}{c}
\partial _{i}\beta _{jk}=\beta _{ji}\beta _{ik}\text{, \ \ \ }i\neq j\neq k,
\\ 
\\ 
\partial _{i}\beta _{ik}=-\sum_{m\neq i,k}\frac{c_{m}-c_{k}}{c_{i}-c_{k}}%
\beta _{mi}\beta _{mk}\text{, \ \ \ \ \ }i\neq k.%
\end{array}%
\end{equation*}%
It easy to see that the consequences of (\ref{gau}) and (\ref{gaus})%
\begin{equation*}
\begin{array}{c}
S_{i}+\frac{1}{2}\sum_{m\neq i}\beta _{mi}^{2}=n_{i}(r^{i}), \\ 
\\ 
c_{i}S_{i}+\frac{1}{2}\sum_{m\neq i}c_{m}\beta _{mi}^{2}=k_{i}(r^{i})%
\end{array}%
\end{equation*}%
for particular case $\eta _{i}(r^{i})=c_{i}=\func{const}$ are the constraints%
\begin{equation*}
\sum_{m\neq i}(c_{m}-c_{i})\beta _{mi}^{2}=2[k_{i}(r^{i})-c_{i}n_{i}(r^{i})].
\end{equation*}%
Thus, if $N=3$, then above system%
\begin{eqnarray*}
(c_{2}-c_{1})\beta _{21}^{2}+(c_{3}-c_{1})\beta _{31}^{2} &=&l_{1}(r^{1}), \\
(c_{1}-c_{2})\beta _{12}^{2}+(c_{3}-c_{2})\beta _{32}^{2} &=&l_{2}(r^{2}), \\
(c_{1}-c_{3})\beta _{13}^{2}+(c_{2}-c_{3})\beta _{23}^{2} &=&l_{3}(r^{3}),
\end{eqnarray*}%
can be parameterized%
\begin{eqnarray*}
\beta _{21} &=&\sqrt{\frac{l_{1}(r^{1})}{c_{2}-c_{1}}}\cosh u\text{, \ \ \ \
\ }\beta _{31}=\sqrt{\frac{l_{1}(r^{1})}{c_{1}-c_{3}}}\sinh u\text{,} \\
\beta _{12} &=&\sqrt{\frac{l_{2}(r^{2})}{c_{1}-c_{2}}}\cosh \upsilon \text{,
\ \ \ \ \ }\beta _{32}=\sqrt{\frac{l_{2}(r^{2})}{c_{2}-c_{3}}}\sinh \upsilon 
\text{,} \\
\beta _{13} &=&\sqrt{\frac{l_{3}(r^{3})}{c_{1}-c_{3}}}\cosh w\text{, \ \ \ \
\ }\beta _{23}=\sqrt{\frac{l_{3}(r^{3})}{c_{3}-c_{2}}}\sinh w\text{.}
\end{eqnarray*}%
Then the $3$-wave system has the form%
\begin{eqnarray*}
\frac{1}{\sqrt{l_{1}(r^{1})}}\partial _{1}w &=&\sqrt{\frac{c_{2}-c_{3}}{%
(c_{1}-c_{2})(c_{1}-c_{3})}}\cosh u\text{, \ \ \ }\frac{1}{\sqrt{l_{3}(r^{3})%
}}\partial _{3}u=\sqrt{\frac{c_{1}-c_{2}}{(c_{1}-c_{3})(c_{2}-c_{3})}}\sinh
w, \\
\frac{1}{\sqrt{l_{1}(r^{1})}}\partial _{1}\upsilon &=&\sqrt{\frac{c_{2}-c_{3}%
}{(c_{1}-c_{2})(c_{1}-c_{3})}}\sinh u\text{, \ \ \ \ }\frac{1}{\sqrt{%
l_{2}(r^{2})}}\partial _{2}u=\sqrt{\frac{c_{1}-c_{3}}{%
(c_{2}-c_{1})(c_{2}-c_{3})}}\sinh \upsilon , \\
\frac{1}{\sqrt{l_{2}(r^{2})}}\partial _{2}w &=&\sqrt{\frac{c_{1}-c_{3}}{%
(c_{1}-c_{2})(c_{3}-c_{2})}}\cosh \upsilon \text{, \ \ \ \ }\frac{1}{\sqrt{%
l_{3}(r^{3})}}\partial _{3}\upsilon =\sqrt{\frac{c_{1}-c_{2}}{%
(c_{1}-c_{3})(c_{2}-c_{3})}}\cosh w.
\end{eqnarray*}%
Introducing new independent variables (scaling Riemann invariants)%
\begin{eqnarray*}
p &=&\sqrt{\frac{c_{2}-c_{3}}{(c_{1}-c_{2})(c_{1}-c_{3})}}\int \sqrt{%
l_{1}(r^{1})}dr^{1}, \\
q &=&i\sqrt{\frac{c_{1}-c_{3}}{(c_{1}-c_{2})(c_{2}-c_{3})}}\int \sqrt{%
l_{2}(r^{2})}dr^{2}, \\
s &=&\sqrt{\frac{c_{1}-c_{2}}{(c_{1}-c_{3})(c_{2}-c_{3})}}\int \sqrt{%
l_{3}(r^{3})}dr^{3},
\end{eqnarray*}%
finally one can obtain the system (see \cite{fer})%
\begin{eqnarray*}
\partial _{p}w &=&\cosh u\text{, \ \ \ }\partial _{s}u=\sinh w, \\
\partial _{p}\upsilon &=&\sinh u\text{, \ \ \ \ }\partial _{q}u=\sinh
\upsilon , \\
\partial _{q}w &=&\cosh \upsilon \text{, \ \ \ }\partial _{s}\upsilon =\cosh
w.
\end{eqnarray*}%
For instance, eliminating field variables $\upsilon $ and $w$ ($\upsilon =%
\func{arcsinh}u_{q}$, \ $w=\func{arcsinh}u_{s}$), one can obtain two
modified $\sinh $-Gordon equations%
\begin{equation*}
u_{pq}=\sinh u\sqrt{u_{q}^{2}+1}\text{, \ \ \ \ }u_{sp}=\cosh u\sqrt{%
u_{s}^{2}+1}
\end{equation*}%
and the degenerated twice modified $\sinh $-Gordon equation (\cite{Zyk})%
\begin{equation*}
u_{sq}=\sqrt{(u_{q}^{2}+1)(u_{s}^{2}+1)}.
\end{equation*}%
Such equations can be obtained as well as with respect to other field $%
\upsilon $ and\ $w$%
\begin{eqnarray*}
\upsilon _{pq} &=&\sinh \upsilon \sqrt{\upsilon _{p}^{2}+1},\ \ \ \ \upsilon
_{sq}=\cosh \upsilon \sqrt{\upsilon _{s}^{2}-1}\text{, \ \ \ \ }\upsilon
_{sp}=\sqrt{(\upsilon _{p}^{2}+1)(\upsilon _{s}^{2}-1)}, \\
w_{sq} &=&\cosh w\sqrt{w_{q}^{2}-1}\text{, \ \ \ }w_{sp}=\sinh w\sqrt{%
w_{p}^{2}-1}\text{, \ \ \ }w_{pq}=\sqrt{(w_{p}^{2}-1)(w_{q}^{2}-1)}.
\end{eqnarray*}%
The substitution%
\begin{equation*}
z=\upsilon +w=w+\ln [w_{q}+\sqrt{w_{q}^{2}-1}]
\end{equation*}%
transforms these equations into%
\begin{equation}
z_{sq}=\sinh z\text{, \ \ \ \ \ }z_{ssp}=z_{s}\sqrt{z_{p}^{2}+z_{sp}^{2}}%
-z_{p}.  \label{sin}
\end{equation}%
The first one is well-known Bonnet equation (the $\sinh $-Gordon equation),
the second one is the next commuting flow from a hierarchy of the potential
modified KdV equation (see \cite{Maks+Kamch})%
\begin{equation*}
z_{\tau }=z_{sss}-\frac{1}{2}z_{s}^{3}.
\end{equation*}%
Thus, a spectral problem for a \textit{triple} of the modified $\sinh $%
-Gordon equations has a form%
\begin{eqnarray*}
\left( 
\begin{array}{c}
\psi \\ 
\bar{\psi} \\ 
\tilde{\psi}%
\end{array}%
\right) _{p} &=&\left( 
\begin{array}{ccc}
0 & (1-\frac{1}{\zeta })\cosh u & \frac{1}{\zeta }\sinh u \\ 
\cosh u & 0 & 0 \\ 
\sinh u & 0 & 0%
\end{array}%
\right) \left( 
\begin{array}{c}
\psi \\ 
\bar{\psi} \\ 
\tilde{\psi}%
\end{array}%
\right) , \\
\left( 
\begin{array}{c}
\psi \\ 
\bar{\psi} \\ 
\tilde{\psi}%
\end{array}%
\right) _{q} &=&\left( 
\begin{array}{ccc}
0 & \cosh \upsilon & 0 \\ 
-\frac{\zeta }{1-\zeta }\cosh \upsilon & 0 & \frac{1}{1-\zeta }\sinh \upsilon
\\ 
0 & \sinh \upsilon & 0%
\end{array}%
\right) \left( 
\begin{array}{c}
\psi \\ 
\bar{\psi} \\ 
\tilde{\psi}%
\end{array}%
\right) , \\
\left( 
\begin{array}{c}
\psi \\ 
\bar{\psi} \\ 
\tilde{\psi}%
\end{array}%
\right) _{s} &=&\left( 
\begin{array}{ccc}
0 & 0 & \cosh w \\ 
0 & 0 & \sinh w \\ 
\zeta \cosh w & (1-\zeta )\sinh w & 0%
\end{array}%
\right) \left( 
\begin{array}{c}
\psi \\ 
\bar{\psi} \\ 
\tilde{\psi}%
\end{array}%
\right) ,
\end{eqnarray*}%
where $\psi _{1}=\sqrt{c_{2}-c_{3}}\psi $, \ $\psi _{2}=\sqrt{c_{3}-c_{1}}%
\bar{\psi}$, \ $\psi _{3}=\sqrt{c_{1}-c_{2}}\tilde{\psi}$, a spectral
parameter is%
\begin{equation*}
\zeta =-\frac{(\lambda +c_{1})(c_{2}-c_{3})}{(\lambda +c_{3})(c_{1}-c_{2})}.
\end{equation*}%
\textbf{Remark}: Each above linear system can be reduced to well-known
spectral problem 2x2 for the modified $\sinh $-Gordon equation (see \cite%
{Zyk}) by transformations described in \cite{Kamch}. This becomes obvious if
take into account that all three above linear systems 3x3 have constraint
(see (\ref{metr}))%
\begin{equation*}
\zeta \psi ^{2}+(1-\zeta )\bar{\psi}^{2}-\tilde{\psi}^{2}=\func{const}.
\end{equation*}%
It means that the functions $\psi $, $\bar{\psi}$, $\tilde{\psi}$ are
\textquotedblleft squares of eigenfunctions\textquotedblright\ for
corresponding spectral transform 2x2 (see \cite{Kamch}). Also, every above
linear system 3x3 can be written as a \textit{scalar} spectral problem of
the third order. For instance, the third linear system is equivalent to the
third order linear equation%
\begin{equation}
\tilde{\psi}_{sss}-\frac{w_{ss}}{w_{s}}\tilde{\psi}_{ss}-(\zeta +\sinh
^{2}w+w_{s}^{2})\tilde{\psi}_{s}+[(\zeta +\sinh ^{2}w)\frac{w_{ss}}{w_{s}}%
-3w_{s}\sinh w\cosh w]\tilde{\psi}=0.  \label{lin}
\end{equation}%
At the same time, the well-known fact is (see for instance, \cite{Oevel})
that the Yajima-Oikawa hierarchy (the Yajima-Oikawa system also is known as
the \textit{long-short resonance}) is associated with the spectral problem%
\begin{equation}
\hat{L}\tilde{\psi}=\zeta \tilde{\psi},  \label{spek}
\end{equation}%
where%
\begin{equation}
\hat{L}=\partial _{s}^{2}-a_{0}+a_{1}\partial _{s}^{-1}a_{2}.  \label{op}
\end{equation}%
Substituting (\ref{op}) in (\ref{spek}) and comparing with (\ref{lin}), one
can obtain expressions for the coefficients $a_{k}$%
\begin{equation*}
a_{0}=\sinh ^{2}w+w_{s}^{2}\text{, \ \ \ \ }a_{1}=w_{s}\text{, \ \ \ \ }%
a_{2}=w_{ss}-\sinh w\cosh w.
\end{equation*}%
That means, that the $\sinh $-Gordon equation is embedded not only to the
KdV hierarchy but also to the Yajima-Oikawa hierarchy. In other words: the
pseudo-differential Manin-Sato operator%
\begin{equation*}
\hat{L}=\partial _{s}+A_{0}\partial _{s}^{-1}+A_{1}\partial _{s}^{-2}+...
\end{equation*}%
associated with the KP hierarchy has finite-component reductions \cite{Orlov}%
\begin{equation*}
\hat{L}^{n}=\partial _{s}^{n}+A_{0,n}\partial _{s}^{n-2}+...+A_{n-2,n}+%
\overset{N}{\sum_{k=1}}B_{k,n}\partial _{s}^{-1}C_{k,n},
\end{equation*}%
where $N$ is an arbitrary natural number. Thus the KP reduction determined
by operator%
\begin{equation*}
\hat{L}=\partial _{s}^{2}-(\sinh ^{2}w+w_{s}^{2})+w_{s}\partial
_{s}^{-1}(w_{ss}-\sinh w\cosh w),
\end{equation*}%
describes three-component bi-Hamiltonian structures of hydrodynamic type
systems.

\textbf{Remark}: One can introduce field variable $c=z_{p}$, then re-scale
independent variables $\partial _{s}\rightarrow \varepsilon \partial _{s}$,
\ $\partial _{p}\rightarrow \varepsilon ^{-1}\partial _{p}$, then a second
equation from (\ref{sin}) has a form%
\begin{equation*}
\varepsilon ^{2}\partial _{s}c=\partial _{p}\frac{c+\varepsilon ^{2}c_{ss}}{%
\sqrt{c^{2}+\varepsilon ^{2}c_{s}^{2}}}.
\end{equation*}%
A limit of this equation with respect to the parameter $\varepsilon $ when $%
\varepsilon \rightarrow 0$ yields%
\begin{equation*}
\partial _{s}c=\partial _{p}[\frac{c_{ss}}{c}-\frac{c_{s}^{2}}{2c^{2}}],
\end{equation*}%
which is nothing but again the $\sinh $-Gordon equation%
\begin{equation*}
b_{sp}=\sinh b,
\end{equation*}%
where $\ln c=b$.

\textbf{Remark}: The substitutions%
\begin{equation*}
z^{1}=w\pm \upsilon \text{, \ \ \ \ }z^{2}=\upsilon \pm u\text{, \ \ \ \ }%
z^{3}=u\pm w
\end{equation*}%
connect solutions of the $\sinh $-Gordon equation%
\begin{equation*}
z_{sq}^{1}=\sinh z^{1}\text{, \ \ \ \ }z_{pq}^{2}=\sinh z^{2}\text{, \ \ \ \ 
}z_{sp}^{3}=\cosh z^{3}.
\end{equation*}

\subsection{General case, $\protect\eta _{i}(r^{i})=r^{i}$}

In general case $\eta _{i}(r^{i})=r^{i}$ the spectral problem

\begin{equation*}
\begin{array}{c}
\partial _{i}\psi _{j}^{(k)}=\beta _{ji}\psi _{i}^{(k)}\text{, \ \ \ \ }%
i\neq j, \\ 
\\ 
0=(\lambda +r^{i})\partial _{i}\psi _{i}^{(k)}+\frac{1}{2}\psi
_{i}^{(k)}+\sum_{m\neq i}(\lambda +r^{m})\beta _{mi}\psi _{m}^{(k)}\text{, \
\ \ \ }k=1\text{, }2\text{, ... , }N,%
\end{array}%
\end{equation*}%
determines the nonlinear system%
\begin{equation*}
\begin{array}{c}
\partial _{i}\beta _{jk}=\beta _{ji}\beta _{ik}\text{, \ \ \ }i\neq j\neq k,
\\ 
\\ 
\partial _{i}\beta _{ik}=\frac{1}{r^{i}-r^{k}}[-\frac{1}{2}(\beta
_{ik}+\beta _{ki})+\sum_{m\neq i,k}(r^{k}-r^{m})\beta _{mi}\beta _{mk}]\text{%
, \ \ \ \ \ }i\neq k.%
\end{array}%
\end{equation*}%
It is easy to see that consequences of (\ref{gau}) and (\ref{gaus}) yield
another \textquotedblleft first integrals\textquotedblright 
\begin{equation*}
\sum_{m\neq i}(r^{i}-r^{m})\beta _{mi}^{2}=V_{i}.
\end{equation*}%
For instance, if $N=3$, then above constraints%
\begin{eqnarray*}
(r^{1}-r^{2})\beta _{21}^{2}+(r^{1}-r^{3})\beta _{31}^{2} &=&V_{1}, \\
(r^{2}-r^{1})\beta _{12}^{2}+(r^{2}-r^{3})\beta _{32}^{2} &=&V_{2}, \\
(r^{3}-r^{1})\beta _{13}^{2}+(r^{3}-r^{2})\beta _{23}^{2} &=&V_{3}.
\end{eqnarray*}%
describe a bi-Hamiltonian structure determined by the flat metrics $g^{ii}$
and $r^{i}g^{ii}$.

Using parametrization%
\begin{eqnarray*}
\beta _{21} &=&\sqrt{\frac{V_{1}}{r^{1}-r^{2}}}\cosh u\text{, \ \ \ \ }\beta
_{31}=\sqrt{\frac{V_{1}}{r^{3}-r^{1}}}\sinh u\text{,} \\
\beta _{32} &=&\sqrt{\frac{V_{2}}{r^{2}-r^{3}}}\cosh \upsilon \text{, \ \ \
\ }\beta _{12}=\sqrt{\frac{V_{2}}{r^{1}-r^{2}}}\sinh \upsilon \text{,} \\
\beta _{13} &=&\sqrt{\frac{V_{3}}{r^{3}-r^{1}}}\cosh w\text{, \ \ \ \ }\beta
_{23}=\sqrt{\frac{V_{3}}{r^{2}-r^{3}}}\sinh w\text{,}
\end{eqnarray*}%
the system (\ref{one}) transforms into

\begin{eqnarray*}
\partial _{1}w &=&-\frac{\sqrt{V_{1}/V_{3}}}{2(r^{3}-r^{1})}\sinh u\sinh w+%
\sqrt{\frac{r^{2}-r^{3}}{(r^{1}-r^{2})(r^{3}-r^{1})}}\sqrt{V_{1}}\cosh u%
\text{,} \\
\partial _{3}u &=&-\frac{\sqrt{V_{3}/V_{1}}}{2(r^{3}-r^{1})}\cosh w\cosh u+%
\sqrt{\frac{r^{1}-r^{2}}{(r^{3}-r^{1})(r^{2}-r^{3})}}\sqrt{V_{3}}\sinh w%
\text{,} \\
V_{13} &=&\frac{\sqrt{V_{1}V_{3}}}{r^{3}-r^{1}}\cosh w\sinh u,
\end{eqnarray*}

\begin{eqnarray*}
\partial _{1}\upsilon  &=&-\frac{\sqrt{V_{1}/V_{2}}}{2(r^{1}-r^{2})}\cosh
u\cosh \upsilon +\sqrt{\frac{r^{2}-r^{3}}{(r^{1}-r^{2})(r^{3}-r^{1})}}\sqrt{%
V_{1}}\sinh u\text{,} \\
\partial _{2}u &=&-\frac{\sqrt{V_{2}/V_{1}}}{2(r^{1}-r^{2})}\sinh \upsilon
\sinh u+\sqrt{\frac{r^{3}-r^{1}}{(r^{1}-r^{2})(r^{2}-r^{3})}}\sqrt{V_{2}}%
\cosh \upsilon \text{,} \\
V_{12} &=&\frac{\sqrt{V_{1}V_{2}}}{r^{1}-r^{2}}\sinh \upsilon \cosh u,
\end{eqnarray*}

\begin{eqnarray*}
\partial _{2}w &=&-\frac{\sqrt{V_{2}/V_{3}}}{2(r^{2}-r^{3})}\cosh \upsilon
\cosh w+\sqrt{\frac{r^{3}-r^{1}}{(r^{1}-r^{2})(r^{2}-r^{3})}}\sqrt{V_{2}}%
\sinh \upsilon \text{,} \\
\partial _{3}\upsilon  &=&-\frac{\sqrt{V_{3}/V_{2}}}{2(r^{2}-r^{3})}\sinh
w\sinh \upsilon +\sqrt{\frac{r^{1}-r^{2}}{(r^{3}-r^{1})(r^{2}-r^{3})}}\sqrt{%
V_{3}}\cosh w\text{,} \\
V_{23} &=&\frac{\sqrt{V_{2}V_{3}}}{r^{2}-r^{3}}\cosh \upsilon \sinh w,
\end{eqnarray*}%
where $V_{ik}=\beta _{ik}\beta _{ki}$. Each above triple of equation depends
from third independent variable as a parameter, which can be eliminated by
shift along both other independent variables. Thus, every above system can
be written in form%
\begin{eqnarray*}
\partial _{x}w &=&\frac{\sqrt{V_{x}/V_{y}}}{2(x-y)}\sinh u\sinh w+\sqrt{%
\frac{y}{x(x-y)}}\sqrt{V_{x}}\cosh u\text{,} \\
\partial _{y}u &=&\frac{\sqrt{V_{y}/V_{x}}}{2(x-y)}\cosh w\cosh u+\sqrt{%
\frac{x}{y(x-y)}}\sqrt{V_{y}}\sinh w\text{,} \\
V_{xy} &=&-\frac{\sqrt{V_{x}V_{y}}}{x-y}\cosh w\sinh u.
\end{eqnarray*}

\section{Co-dimension 1}

In previous section a flat case is considered, where relationship between
two conjugate linear problems (\ref{a}), (\ref{b}) is determined by the
linear differential operator of the first order (\ref{c}). In general case
of arbitrary curvature matric (\ref{met}) such relationship becomes nonlocal
(see for details \cite{Ferr})%
\begin{equation}
H_{i}=\partial _{i}\psi _{i}+\sum_{m\neq i}\beta _{mi}\psi _{m}+\overset{M}{%
\sum_{k=1}}\overset{M}{\sum_{n=1}}\varepsilon _{kn}H_{i}^{(k)}a^{(n)},
\label{non}
\end{equation}%
where $\partial _{i}a^{(n)}=\psi _{i}H_{i}^{(n)}$, $\varepsilon _{kn}$ are
constant symmetric non-degenerate matrix, and a number of particular
solutions $H_{i}^{(n)}$ of (\ref{a}) is determined by the Gauss equation
(see \cite{Ferr})%
\begin{equation*}
\partial _{i}\beta _{ij}+\partial _{j}\beta _{ji}+\sum_{m\neq i,j}\beta
_{mi}\beta _{mj}=\overset{M}{\sum_{k=1}}\overset{M}{\sum_{n=1}}\varepsilon
_{kn}H_{i}^{(k)}H_{j}^{(n)}\text{, \ \ \ \ }i\neq j.
\end{equation*}%
Thus, in general case the description of pairs of nonlocal differential
operators (\ref{non}) (similar to the flat case) is equivalent to the system%
\begin{equation*}
\begin{array}{c}
\partial _{i}\beta _{jk}=\beta _{ji}\beta _{ik}\text{, \ \ \ }i\neq j\neq k,
\\ 
\partial _{i}\beta _{ij}+\partial _{j}\beta _{ji}+\sum_{m\neq i,j}\beta
_{mi}\beta _{mj}=\overset{M}{\sum_{k=1}}\overset{M}{\sum_{n=1}}\varepsilon
_{kn}H_{i}^{(k)}H_{j}^{(n)}\text{, \ \ \ \ }i\neq j, \\ 
r^{i}\partial _{i}\beta _{ij}+r^{j}\partial _{j}\beta _{ji}+\frac{1}{2}\beta
_{ij}+\frac{1}{2}\beta _{ji}+\sum_{m\neq i,j}r^{m}\beta _{mi}\beta _{mj}=%
\overset{\tilde{M}}{\sum_{k=1}}\overset{\tilde{M}}{\sum_{n=1}}\tilde{%
\varepsilon}_{kn}\tilde{H}_{i}^{(k)}\tilde{H}_{j}^{(n)}\text{, \ \ \ \ }%
i\neq j.%
\end{array}%
\end{equation*}%
In case of non-flat metric $\eta _{i}(r^{i})g^{ii}$ the Gauss equation%
\begin{equation*}
\eta _{i}(r^{i})\partial _{i}\beta _{ij}+\eta _{j}(r^{j})\partial _{j}\beta
_{ji}+\frac{1}{2}\eta _{i}^{\prime }(r^{i})\beta _{ij}+\frac{1}{2}\eta
_{j}^{\prime }(r^{j})\beta _{ij}+\sum_{m\neq i,j}\eta _{m}(r^{m})\beta
_{mi}\beta _{mj}=\overset{M}{\sum_{k=1}}\overset{M}{\sum_{n=1}}\varepsilon
_{kn}H_{i}^{(k)}H_{j}^{(n)}\text{, \ \ \ \ }i\neq k
\end{equation*}%
has $N$ \textquotedblleft first integrals\textquotedblright 
\begin{equation*}
\eta _{i}S_{i}+\frac{1}{2}\eta _{i}^{\prime }V_{i}+\frac{1}{2}\sum_{m\neq
i}\eta _{m}\beta _{mi}^{2}=\frac{1}{2}\overset{M}{\sum_{k=1}}\overset{M}{%
\sum_{n=1}}\varepsilon _{kn}H_{i}^{(k)}H_{i}^{(n)}+k_{i}(r^{i}),
\end{equation*}%
where $k_{i}(r^{i})$ are some functions.

In this section we restrict our consideration on two simplest cases: a flat
metric $g^{ii}$ and a metric $\tilde{g}^{ii}$ of co-dimension 1, when $%
\tilde{g}^{ii}=c_{i}g^{ii}$ or $\tilde{g}^{ii}=r^{i}g^{ii}$. In the first
case the system 
\begin{equation}
\begin{array}{c}
\partial _{i}\beta _{jk}=\beta _{ji}\beta _{ik}\text{, \ \ \ }i\neq j\neq k,
\\ 
\\ 
\partial _{i}\beta _{ij}+\partial _{j}\beta _{ji}+\sum_{m\neq i,j}\beta
_{mi}\beta _{mj}=0\text{, \ \ \ \ }i\neq j, \\ 
\\ 
c_{i}\partial _{i}\beta _{ij}+c_{j}\partial _{j}\beta _{ji}+\sum_{m\neq
i,j}c_{m}\beta _{mi}\beta _{mj}=H_{i}H_{j}\text{, \ \ \ \ }i\neq j,%
\end{array}
\label{first}
\end{equation}%
is a result of a compatibility condition of the linear system (see \cite%
{Mokh})

\begin{equation*}
\begin{array}{c}
\partial _{i}\psi _{j}=\beta _{ji}\psi _{i}\text{, \ \ \ \ }i\neq j, \\ 
\\ 
0=(\lambda +c_{i})\partial _{i}\psi _{i}+\sum_{m\neq i}(\lambda +c_{m})\beta
_{mi}\psi _{m}+H_{i}a, \\ 
\\ 
\partial _{i}a=H_{i}\psi _{i}.%
\end{array}%
\end{equation*}%
In the second case the system%
\begin{equation}
\begin{array}{c}
\partial _{i}\beta _{jk}=\beta _{ji}\beta _{ik}\text{, \ \ \ }i\neq j\neq k,
\\ 
\\ 
\partial _{i}\beta _{ij}+\partial _{j}\beta _{ji}+\sum_{m\neq i,j}\beta
_{mi}\beta _{mj}=0\text{, \ \ \ \ }i\neq j, \\ 
\\ 
r^{i}\partial _{i}\beta _{ij}+r^{j}\partial _{j}\beta _{ji}+\frac{1}{2}\beta
_{ij}+\frac{1}{2}\beta _{ji}+\sum_{m\neq i,j}r^{m}\beta _{mi}\beta
_{mj}=H_{i}H_{j}\text{, \ \ \ \ }i\neq j,%
\end{array}
\label{sec}
\end{equation}%
is a result of a compatibility condition of the linear system (see \cite%
{Mokh})

\begin{equation*}
\begin{array}{c}
\partial _{i}\psi _{j}=\beta _{ji}\psi _{i}\text{, \ \ \ \ }i\neq j, \\ 
\\ 
0=(\lambda +r^{i})\partial _{i}\psi _{i}+\frac{1}{2}\psi _{i}+\sum_{m\neq
i}(\lambda +r^{m})\beta _{mi}\psi _{m}+H_{i}a, \\ 
\\ 
\partial _{i}a=H_{i}\psi _{i}.%
\end{array}%
\end{equation*}

\subsection{Particular case, $\protect\eta _{i}(r^{i})=c_{i}=\func{const}$}

As well as in the flat case the nonlinear system (\ref{first})%
\begin{equation*}
\begin{array}{c}
\partial _{i}\beta _{jk}=\beta _{ji}\beta _{ik}\text{, \ \ \ }i\neq j\neq k,
\\ 
\\ 
\partial _{i}\beta _{ik}=\frac{1}{c_{i}-c_{k}}[H_{i}H_{k}+\sum_{m\neq
i,k}(c_{k}-c_{m})\beta _{mi}\beta _{mk}]\text{, \ \ \ \ \ }i\neq k%
\end{array}%
\end{equation*}%
has the constraints%
\begin{equation*}
\begin{array}{c}
S_{i}+\frac{1}{2}\sum_{m\neq i}\beta _{mi}^{2}=n_{i}(r^{i}), \\ 
\\ 
c_{i}S_{i}+\frac{1}{2}\sum_{m\neq i}c_{m}\beta _{mi}^{2}=\frac{1}{2}%
H_{i}^{2}+k_{i}(r^{i}).%
\end{array}%
\end{equation*}%
Thus, one has%
\begin{equation*}
\sum_{m\neq i}(c_{i}-c_{m})\beta
_{mi}^{2}+H_{i}^{2}=2[c_{i}n_{i}(r^{i})-k_{i}(r^{i})].
\end{equation*}%
When $N=3$, then above system has a form%
\begin{eqnarray*}
(c_{1}-c_{2})\beta _{21}^{2}+(c_{1}-c_{3})\beta _{31}^{2}+H_{1}^{2}
&=&l_{1}(r^{1}), \\
(c_{2}-c_{1})\beta _{12}^{2}+(c_{2}-c_{3})\beta _{32}^{2}+H_{2}^{2}
&=&l_{2}(r^{2}), \\
(c_{3}-c_{1})\beta _{13}^{2}+(c_{3}-c_{2})\beta _{23}^{2}+H_{3}^{2}
&=&l_{3}(r^{3}).
\end{eqnarray*}%
Introducing new independent variables (re-scaling Riemann invariants)%
\begin{eqnarray*}
p &=&i\sqrt{\frac{c_{2}-c_{3}}{(c_{1}-c_{2})(c_{1}-c_{3})}}\int \sqrt{%
l_{1}(r^{1})}dr^{1}, \\
q &=&\sqrt{\frac{c_{1}-c_{3}}{(c_{1}-c_{2})(c_{2}-c_{3})}}\int \sqrt{%
l_{2}(r^{2})}dr^{2}, \\
s &=&\sqrt{\frac{c_{1}-c_{2}}{(c_{1}-c_{3})(c_{2}-c_{3})}}\int \sqrt{%
l_{3}(r^{3})}dr^{3},
\end{eqnarray*}%
and new dependent functions%
\begin{eqnarray*}
H_{1} &=&\sqrt{l_{1}}R_{1}\text{, \ \ }H_{2}=\sqrt{l_{2}}S_{2}\text{, \ \ }%
H_{3}=\sqrt{l_{3}}P_{3}, \\
\beta _{13} &=&i\sqrt{\frac{l_{3}}{c_{1}-c_{3}}}P_{2}\text{, \ \ \ }\beta
_{31}=\sqrt{\frac{l_{1}}{c_{1}-c_{3}}}R_{2}\text{, \ \ \ \ }\beta _{21}=%
\sqrt{\frac{l_{1}}{c_{1}-c_{2}}}R_{3}, \\
\beta _{23} &=&-i\sqrt{\frac{l_{3}}{c_{2}-c_{3}}}P_{1}\text{, \ \ \ }\beta
_{12}=i\sqrt{\frac{l_{2}}{c_{1}-c_{2}}}S_{3}\text{, \ \ \ \ }\beta _{32}=%
\sqrt{\frac{l_{2}}{c_{2}-c_{3}}}S_{1},
\end{eqnarray*}%
the system (\ref{one}) can be written in form pairwise commuting hyperbolic
systems%
\begin{eqnarray*}
\partial _{p}P_{1} &=&iR_{3}P_{2}\text{, \ \ \ \ \ \ \ \ \ \ \ \ \ \ \ \ \ \
\ \ \ \ \ \ \ \ \ \ \ \ \ \ \ \ \ \ }\partial _{s}R_{1}=\frac{\Delta }{\sqrt{%
1-\Delta ^{2}}}P_{3}R_{2}, \\
\partial _{p}P_{2} &=&-iP_{1}R_{3}-\frac{\sqrt{1-\Delta ^{2}}}{\Delta }%
P_{3}R_{1}\text{, \ \ \ \ \ \ \ \ \ \ \ \ }\partial _{s}R_{2}=iR_{3}P_{1}-%
\frac{\Delta }{\sqrt{1-\Delta ^{2}}}R_{1}P_{3}, \\
\partial _{p}P_{3} &=&\frac{\sqrt{1-\Delta ^{2}}}{\Delta }R_{1}P_{2}\text{,
\ \ \ \ \ \ \ \ \ \ \ \ \ \ \ \ \ \ \ \ \ \ \ \ \ }\partial
_{s}R_{3}=-iP_{1}R_{2}, \\
&& \\
\partial _{p}S_{1} &=&R_{2}S_{3}\text{, \ \ \ \ \ \ \ \ \ \ \ \ \ \ \ \ \ \
\ \ \ \ \ \ \ \ \ \ \ \ \ \ \ \ \ \ \ }\partial _{q}R_{1}=\Delta S_{2}R_{3},
\\
\partial _{p}S_{2} &=&\frac{1}{\Delta }R_{1}S_{3}\text{, \ \ \ \ \ \ \ \ \ \
\ \ \ \ \ \ \ \ \ \ \ \ \ \ \ \ \ \ \ \ \ \ \ \ }\partial
_{q}R_{2}=S_{1}R_{3}, \\
\partial _{p}S_{3} &=&-R_{2}S_{1}-\frac{1}{\Delta }R_{1}S_{2}\text{, \ \ \ \
\ \ \ \ \ \ \ \ \ \ \ \ \ \ \ \ \ \ }\partial _{q}R_{3}=-S_{1}R_{2}-\Delta
S_{2}R_{1}, \\
&& \\
\partial _{q}P_{1} &=&iS_{3}P_{2}+i\sqrt{1-\Delta ^{2}}P_{3}S_{2}\text{, \ \
\ \ \ \ \ \ \ \ \ \ \ }\partial _{s}S_{1}=-S_{3}P_{2}-\frac{1}{\sqrt{%
1-\Delta ^{2}}}S_{2}P_{3}, \\
\partial _{q}P_{2} &=&-iS_{3}P_{1}\text{, \ \ \ \ \ \ \ \ \ \ \ \ \ \ \ \ \
\ \ \ \ \ \ \ \ \ \ \ \ \ \ \ \ \ }\partial _{s}S_{2}=\frac{1}{\sqrt{%
1-\Delta ^{2}}}P_{3}S_{1}, \\
\partial _{q}P_{3} &=&-i\sqrt{1-\Delta ^{2}}S_{2}P_{1}\text{, \ \ \ \ \ \ \
\ \ \ \ \ \ \ \ \ \ \ \ \ \ \ }\partial _{s}S_{3}=P_{2}S_{1}.
\end{eqnarray*}%
where%
\begin{equation*}
\Delta ^{2}=\frac{c_{2}-c_{3}}{c_{1}-c_{3}}\text{, \ \ \ \ }%
P_{1}^{2}+P_{2}^{2}+P_{3}^{2}=1\text{, \ \ \ \ }%
R_{1}^{2}+R_{2}^{2}+R_{3}^{2}=1\text{, \ \ \ \ }%
S_{1}^{2}+S_{2}^{2}+S_{3}^{2}=1.
\end{equation*}%
Each above system is nothing but a some reduction of the famous Cherednik
model of chiral fields related with hierarchy of the Heisenberg magnet (see 
\cite{Chered}).

\subsection{General case, $\protect\eta _{i}(r^{i})=r^{i}$}

In general case the nonlinear system (\ref{sec})%
\begin{equation}
\begin{array}{c}
\partial _{i}\beta _{jk}=\beta _{ji}\beta _{ik}\text{, \ \ \ }i\neq j\neq k,
\\ 
\\ 
\partial _{i}\beta _{ik}=\frac{1}{r^{i}-r^{k}}[H_{i}H_{k}-\frac{1}{2}(\beta
_{ik}+\beta _{ki})+\sum_{m\neq i,k}(r^{k}-r^{m})\beta _{mi}\beta _{mk}]\text{%
, \ \ \ \ \ }i\neq k.%
\end{array}
\label{0}
\end{equation}%
has a set of \textquotedblleft first integrals\textquotedblright 
\begin{equation*}
\begin{array}{c}
S_{i}+\frac{1}{2}\sum_{m\neq i}\beta _{mi}^{2}=n_{i}(r^{i}), \\ 
\\ 
r^{i}S_{i}+\frac{1}{2}V_{i}+\frac{1}{2}\sum_{m\neq i}r^{m}\beta _{mi}^{2}=%
\frac{1}{2}H_{i}^{2}+k_{i}(r^{i}).%
\end{array}%
\end{equation*}%
Then we have%
\begin{equation*}
V_{i}+\sum_{m\neq i}(r^{m}-r^{i})\beta
_{mi}^{2}=H_{i}^{2}+2[k_{i}(r^{i})-c_{i}n_{i}(r^{i})].
\end{equation*}%
Thus, if $N=3$, then the system of $18$ equations (\ref{0}), written in form
of three pairwise commuting hyperbolic flows%
\begin{eqnarray*}
&&%
\begin{array}{ccccc}
\partial _{1}\beta _{23}=\beta _{21}\beta _{13}, &  &  &  & \partial
_{3}\beta _{21}=\beta _{23}\beta _{31}, \\ 
\partial _{1}H_{3}=\beta _{13}H_{1}, &  &  &  & \partial _{3}H_{1}=\beta
_{31}H_{3}, \\ 
\partial _{1}\beta _{13}=\frac{H_{1}H_{3}}{r^{1}-r^{3}}-\frac{\beta
_{13}+\beta _{31}}{2(r^{1}-r^{3})}-\frac{r^{2}-r^{3}}{r^{1}-r^{3}}\beta
_{21}\beta _{23}, &  &  &  & \partial _{3}\beta _{31}=-\frac{H_{1}H_{3}}{%
r^{1}-r^{3}}+\frac{\beta _{13}+\beta _{31}}{2(r^{1}-r^{3})}-\frac{r^{1}-r^{2}%
}{r^{1}-r^{3}}\beta _{21}\beta _{23},%
\end{array}
\\
&&%
\begin{array}{ccccc}
\partial _{1}\beta _{32}=\beta _{31}\beta _{12}, &  &  &  & \partial
_{2}\beta _{31}=\beta _{32}\beta _{21}, \\ 
\partial _{1}H_{2}=\beta _{12}H_{1}, &  &  &  & \partial _{2}H_{1}=\beta
_{21}H_{2}, \\ 
\partial _{1}\beta _{12}=\frac{H_{1}H_{2}}{r^{1}-r^{2}}-\frac{\beta
_{12}+\beta _{21}}{2(r^{1}-r^{2})}+\frac{r^{2}-r^{3}}{r^{1}-r^{2}}\beta
_{31}\beta _{32}, &  &  &  & \partial _{2}\beta _{21}=-\frac{H_{1}H_{2}}{%
r^{1}-r^{2}}+\frac{\beta _{12}+\beta _{21}}{2(r^{1}-r^{2})}-\frac{r^{1}-r^{3}%
}{r^{1}-r^{2}}\beta _{31}\beta _{32},%
\end{array}
\\
&&%
\begin{array}{ccccc}
\partial _{2}\beta _{13}=\beta _{12}\beta _{23}, &  &  &  & \partial
_{3}\beta _{12}=\beta _{13}\beta _{32}, \\ 
\partial _{2}H_{3}=\beta _{23}H_{2}, &  &  &  & \partial _{3}H_{2}=\beta
_{32}H_{3}, \\ 
\partial _{2}\beta _{23}=\frac{H_{2}H_{3}}{r^{2}-r^{3}}-\frac{\beta
_{23}+\beta _{32}}{2(r^{2}-r^{3})}-\frac{r^{1}-r^{3}}{r^{2}-r^{3}}\beta
_{12}\beta _{13}, &  &  &  & \partial _{3}\beta _{32}=-\frac{H_{2}H_{3}}{%
r^{2}-r^{3}}+\frac{\beta _{23}+\beta _{32}}{2(r^{2}-r^{3})}+\frac{r^{1}-r^{2}%
}{r^{2}-r^{3}}\beta _{12}\beta _{13},%
\end{array}%
\end{eqnarray*}%
has three \textquotedblleft first integrals\textquotedblright 
\begin{eqnarray*}
(r^{1}-r^{2})\beta _{21}^{2}+(r^{1}-r^{3})\beta _{31}^{2}+H_{1}^{2} &=&V_{1},
\\
(r^{2}-r^{1})\beta _{12}^{2}+(r^{2}-r^{3})\beta _{32}^{2}+H_{2}^{2} &=&V_{2},
\\
(r^{3}-r^{1})\beta _{13}^{2}+(r^{3}-r^{2})\beta _{23}^{2}+H_{3}^{2} &=&V_{3}.
\end{eqnarray*}%
Since this example and for higher dimensions and co-dimensions presence of
\textquotedblleft the first integrals\textquotedblright\ any more does not
lead to an essential reduction of number of the equations in each of such
hyperbolic systems. For example, at $N=3$ from 18 equations in case of two
flat metrics where ratios of corresponding diagonal coefficients are various
constants, - there are 6 equations of the first order; if ratios of
corresponding diagonal coefficients are some functions of Riemann
invariants, -- there are 9 equations of the first order; if one of metrics
is flat, and the second one is of co-dimension 1, and ratios of
corresponding diagonal coefficients are various constants, - there are 12
equations of the first order; and at last, if ratios of corresponding
diagonal coefficients are some functions, - there are 15 equations of the
first order (see explanation below), - in all other more complicated cases
(any dimension of space $N$ and any co-dimension $M$) the number of
equations in system in involution will be more the of use of
\textquotedblleft first integrals\textquotedblright , than in initial system
of the equations. For example, in last above-stated example - the system has
18 equations. Use of 3 constraints on a first step decreases number of the
equations up to 12. However, as constraints contain the first derivative of
scalar potential $V$ of rotation coefficients of conjugate curvilinear
coordinate nets, it is necessary to add 3 more equations expressing second
derivative from $V$ through rotation coefficients of conjugate curvilinear
coordinate nets. Thus, actually already in this example it is visible, that
above system contains the same number of equations (as any equation of the
second order is equivalent to a pair of the first order equations).

\section{Conclusion}

In this paper it has been shown, that the problem of classification of
semi-Hamiltonian systems of hydrodynamical type with such additional and
important properties as a flatness of the diagonal metrics or metric's
curvature of co-dimension 1 is reduced to the solution of such well-known
integrable systems (by inverse scattering transform) as a negative flow of
the Landau-Lifshitz equation (i.e. the Cherednik model) and generalization
of the Ernst equation well known in Gravity and General Relativity (here a
spectral problem depends explicitly of independent variable).

\section*{Acknowledements}

The author expresses sincere gratitude to E.V. Ferapontov, V.V. Sokolov and
A.M. Kamchatnov for numerous and stimulating discussions, alive interest and
successful advices.


\begin{thebibliography}{99}
\bibitem{Zyk} \emph{A.B. Borisov, S.A. Zykov}, \newblock The dressing chain
of discrete symmetries and the proliferation of nonlinear equations. Theor.
Math. Phys. \textbf{115}, No. 2 (1998) 530--541.

\bibitem{Chered} \emph{I.V. Cherednik}, \newblock On the integrability of
the two-dimensional asymmetric chiral O(3)-field and its quantum analogue.
Yadernaya Fizika (in Russian) Soviet Journal of Nuclear Physics \textbf{33}
(1981) 278-282.

\bibitem{Darboux} \emph{G. Darboux,} \newblock Le\c{c}ons sur les syst\`{e}%
mes orthogonaux et les coordonn\'{e}es curvilignes. Paris, Gautier-Villar,
1910.

\bibitem{Dubr} \emph{B.A. Dubrovin}, \newblock Geometry of 2D topological
field theories, Lecture Notes in Math. \textbf{1620} Springer-Verlag (1996)
120-348.

\bibitem{fer} \emph{E.V. Ferapontov}, \newblock Compatible Poisson brackets
of hydrodynamic type, J. Phys. A: Math. Gen. \textbf{34} (2001) 1-12.

\bibitem{viniti} \emph{E.V. Ferapontov}, \newblock Hamiltonian systems of
hydrodynamic type and their realization on hypersurfaces of a
pseudoeuclidean space, Geom. Sbornik, VINITI \textbf{22} (1990) 59-96;
(English transl. in Soviet J. of Math., \textbf{55} (1991) 1970-1995).

\bibitem{Maks+Fer} \emph{E.V. Ferapontov, M.V. Pavlov}, \newblock %
Quasiclassical limit of Coupled KdV equations. Riemann invariants and
multi-Hamiltonian structure. Physica D \textbf{52} (1991), 211--219.

\bibitem{Oevel} \emph{W. Oevel, W. Strampp}, Constrained KP hierarchy and
Bi-Hamiltonian structures. Comm. Math. Phys. \textbf{157} (1993), 51-81.

\bibitem{Orlov} \emph{A.Yu. Orlov}, \newblock Volterra operator algebra for
zero curvature representation. Universality of KP. Proceedings of the III
Potsdam-Kiev Workshop at Clarkson University, Potsdam, NY, USA, August,
1-11, 1991.

\bibitem{Novik} \emph{B.A. Dubrovin, S.P. Novikov}, \newblock Hydrodynamics
of soliton lattices. \textit{Soviet Scientific Reviews}, Section C: \textit{%
Mathematical Physics Reviews} \textbf{9} Part 4. Harwood Academic Publishers
GmbH, Yverdon (1993) 136 pp.

\bibitem{Ferr} \emph{E.V. Ferapontov}, \newblock Nonlocal Hamiltonian
operators of hydrodynamic type: Differential geometry and Applications,
Amer. Math. Soc. Transl. (2) \textbf{170} (1995) 33-58.

\bibitem{Pavlov} \emph{M.V. Pavlov,} \newblock Elliptic coordinates and
multi-Hamiltonian structures of hydrodynamic type systems. Russian Acad.
Sci. Dokl. Math. \textbf{50} No. 3 (1995) 374-377. \emph{M.V. Pavlov,} %
\newblock Multi-Hamiltonian structures of the Whitham equations. Russian
Acad. Sci. Dokl. Math. \textbf{50} No. 2 (1995) 220-223.

\bibitem{Maks+Tsar} \emph{M.V. Pavlov, S.P. Tsarev,} \newblock Three
Hamiltonian structures of the Egorov hydrodynamic type systems. Funct. Anal.
Appl. \textbf{37} No. 1 (2003) 32-45.

\bibitem{Tsar} \emph{S.P. Tsarev,} \newblock On Poisson brackets and
one-dimensional Hamiltonian systems of hydrodynamic type, Soviet Math. Dokl. 
\textbf{31} (1985) 488--491. \emph{S.P. Tsarev,} \newblock The geometry of
Hamiltonian systems of hydrodynamic type. The generalized hodograph
transform, Math. USSR Izv. \textbf{37} (1991) 397-419.

\bibitem{Maks+Kamch} \emph{A.M. Kamchatnov, M.V. Pavlov}, \newblock On
generating functions in the AKNS hierarchy. Physics Letters A. \textbf{301}
No. 3-4 (2002) 269-274.

\bibitem{Mokh} \emph{O.I. Mokhov}, \newblock Lax pairs for compatible
nonlocal Hamiltonian operators of hydrodynamic type. (Russian) Uspekhi Mat.
Nauk \textbf{57}, No. 6 (2002) 189--190. \emph{O.I. Mokhov}, Compatible
metrics of constant Riemannian curvature: local geometry, nonlinear
equations, and integrability. (Russian) Funktsional. Anal. i Prilozhen. 
\textbf{36} (2002), No. 3, 36--47, 96; translation in Funct. Anal. Appl. 
\textbf{36} (2002), No. 3, 196--204. \emph{O.I. Mokhov}, A Lax pair for
nonsingular bundles of metrics of constant Riemannian curvature. (Russian)
Uspekhi Mat. Nauk \textbf{57} (2002), No. 3(345), 155--156; translation in
Russian Math. Surveys \textbf{57} (2002), No. 3, 603--605.

\bibitem{Kamch} \emph{M.A. Kamchatnov, R.A. Kraenkel}, On the relationship
between a 2x2 matrix and second order scalar spectral problems. J. Phys. A:
Math. Gen. \textbf{35} (2002) L13-L18.
\end{thebibliography}
\end{document}